\documentclass[11pt]{article}
\usepackage[english]{babel}

\usepackage[a4paper,top=2cm,bottom=2cm,left=2.5cm,right=2.5cm,marginparwidth=1.75cm]{geometry}
\usepackage{setspace}
\usepackage{amsmath}
\usepackage{graphicx}
\usepackage{caption}
\usepackage[colorlinks=true, allcolors=blue]{hyperref}

\date{\today}

\usepackage{etoolbox}   

\makeatletter
\patchcmd{\thebibliography}
  {\list}
  {\small
   \list}
  {}{}

\patchcmd{\thebibliography}
  {\setlength{\itemsep}{\z@}} 
  {\setlength{\itemsep}{0pt}%
   \setlength{\parsep}{0pt}}
  {}{}
\makeatother

\begin{document}

\begin{titlepage}
    \centering

    {\Large ESO Expanding Horizons Call \par}
    \vspace{0.8cm}

    {\LARGE\bfseries Binarity Beyond \emph{Gaia}\par}
    \vspace{0.3cm}
    {\large\bfseries The case for a dedicated spectroscopic survey of binary stars\par}
    \vspace{1.5cm}

    {\large
    Borja Anguiano\par}
    \vspace{0.4cm}

    {\small
    Centro de Estudios de F\'isica del Cosmos de Arag\'on (CEFCA),\\
    Plaza San Juan 1, 44001 Teruel, Spain\par}
    \vspace{0.8cm}

    {\small
    E-mail: \texttt{banguiano@cefca.es}
    \par}

    \vfill

    {\small White paper submitted to the ESO ``Expanding Horizons'' Call\par}
    \vspace{0.5cm}

    {\large \today\par}
\end{titlepage}

\setcounter{page}{1}


\begin{abstract}
Stellar multiplicity is a fundamental ingredient of stellar astrophysics, yet binary statistics across the Galaxy remain poorly constrained. The \emph{Gaia} mission has revolutionised binary star astrophysics by delivering high-precision astrometry, photometry and global radial velocities, and by providing hundreds of thousands of non-single-star solutions in DR3. However, the RVS magnitude limit, mission time span and scanning law impose strong selection effects in period, mass ratio, inclination and semi-amplitude, leaving large regions of the binary parameter space either sparsely sampled or effectively inaccessible. In this white paper we outline the case for a dedicated, wide-field, multi-epoch spectroscopic survey explicitly optimised for binary science: deeper than the \emph{Gaia} RVS limit, with flexible cadence from hours to years, and with moderate to high spectral resolution. Using a simplified forward model of \emph{Gaia} DR5-like performance, we highlight the populations for which robust orbital solutions will be rare (ultra short period, very long period, low-amplitude and compact-object binaries), and show how a ``Binarity Beyond \emph{Gaia}'' survey would fill these gaps. Such a programme would deliver a bias correctable census of stellar multiplicity across the Milky Way and provide the spectroscopic backbone needed to exploit binary samples from \emph{Rubin}/LSST, \emph{Roman} and \emph{LISA}.
\end{abstract}

\section{Motivation}

Stellar multiplicity is the rule, not the exception: a large fraction of stars form and evolve in binaries or higher-order systems across the Hertzsprung--Russell diagram~[1,2]. Both single star and binary evolution shape the demographics of white dwarfs, neutron stars and stellar black holes, but binary interaction open additional pathways to Type~Ia supernovae and compact-object mergers. Unresolved binaries also bias colour--magnitude diagrams, integrated-light spectra and chemodynamical inferences, so robust binary statistics as a function of stellar type, chemistry and environment are essential to model the Galaxy and to test binary evolution physics. The \emph{Gaia} mission has transformed binary-star astrophysics by delivering high-precision astrometry, broad-band photometry and global radial velocities for more than a billion stars~[3,4]. The DR3 Non-Single Star (NSS) catalogue provides hundreds of thousands of astrometric, spectroscopic and eclipsing-binary solutions, together with an all-sky sample of eclipsing candidates with light-curve models~[5,6]. These products underpin the first truly global studies of nearby multiplicity, revealing structure in the period, eccentricity and mass-ratio distributions, and enabling population studies of how binarity depends on stellar mass, age and environment~[7,8]. Yet \emph{Gaia}'s spectroscopic view is intrinsically incomplete: the RVS magnitude limit excludes most low-mass dwarfs, white dwarfs and faint halo stars, while the scanning law and finite mission duration impose sparse, non-uniform sampling that strongly biases sensitivity in orbital period, inclination and RV semi-amplitude. Similar selection and cadence effects limit the completeness of \emph{Gaia}'s eclipsing and astrometric binary samples, introducing additional biases in period, inclination and mass ratio. In the post-\emph{Gaia} era, the challenge is therefore not simply to find more binaries, but to turn \emph{Gaia}'s unprecedented yet intrinsically patchy view into a quantitative, bias correctable map of stellar multiplicity. This requires a spectroscopic survey explicitly designed for binary science: deeper than the RVS limit, with multi-epoch radial velocities and cadences optimised from hours to years, and with wide Galactic coverage. Such a survey would also provide the missing electromagnetic context for compact-object binaries in the era of gravitational-wave astronomy: space based observatories such as \emph{LISA}~[11] will detect large populations of short-period double white dwarfs and other compact systems, but their formation channels, merger rates and contribution to the Galactic gravitational-wave background can only be fully understood if the underlying stellar multiplicity is mapped and calibrated across the Milky Way. In parallel, time-domain surveys such as the \emph{Rubin}/\emph{LSST} observatory~[9] and wide-field imaging missions like \emph{Roman}~[10] will uncover vast numbers of eclipsing, transiting and microlensing binaries; without systematic, multi-epoch spectroscopy, however, these samples will lack precise orbital solutions, mass ratios and systemic velocities. A dedicated ``Binarity Beyond \emph{Gaia}'' programme would fill the gaps in period, mass ratio and eccentricity left by \emph{Gaia}, provide the spectroscopic backbone needed to fully exploit \emph{LSST} and \emph{Roman} binary samples, and deliver the homogeneous binary census required for next generation studies of Galactic structure, stellar evolution and multi-messenger astrophysics.

\begin{figure}[!htbp]
\centering
\includegraphics[width=1.\textwidth]{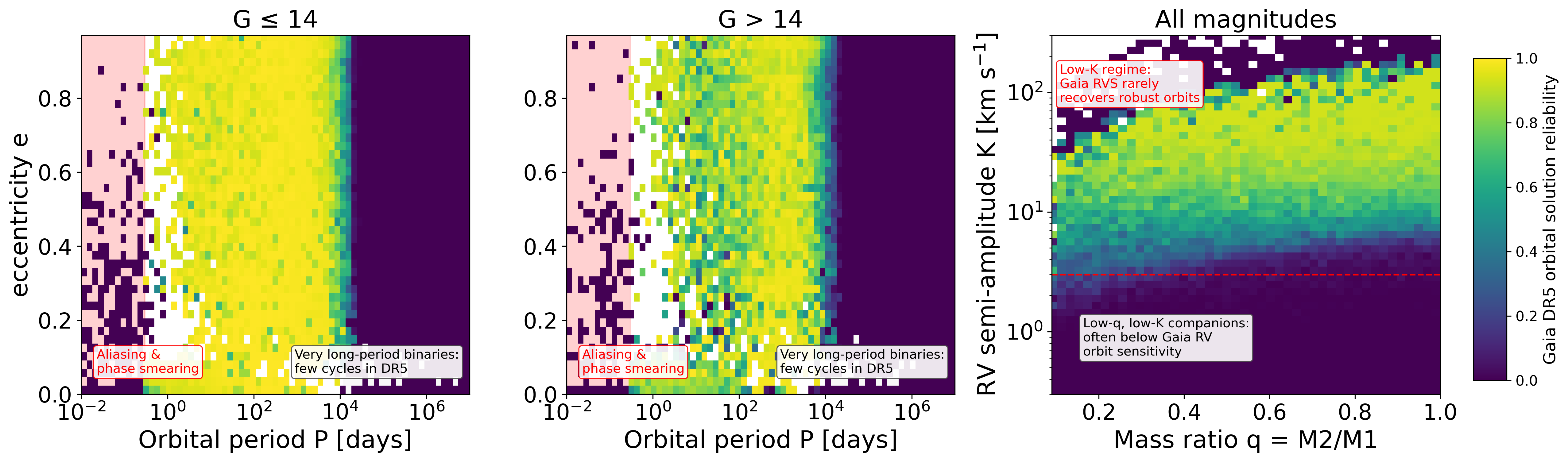}
\caption{\label{fig:wst_galatea_airmass}
Gaia-DR5--like orbital completeness in a toy Milky Way binary population. We draw $\sim 8\times10^{4}$ binaries from a smooth, Raghavan-like period--eccentricity distribution with random masses and inclinations, and assign toy heliocentric distances (100--4000~pc) to model \emph{Gaia}'s astrometric signal. We then propagate them through a simplified Gaia model including RVS and astrometric errors, number of visits and mission baseline. The left and middle panels show, for bright ($G\le 14$) and faint ($G>14$) stars, the mean reliability weight of the recovered orbital solution in the $(P,e)$ plane, combining RV and astrometry; shaded regions highlight ultra-short periods (aliasing and phase smearing) and very long periods with $N_{\rm cycles}\ll1$ in DR5.
The right panel shows the corresponding RV-only reliability in the $(q,K)$ plane, emphasising that low-$q$, low-$K$ systems lie below \emph{Gaia}'s sensitivity for robust spectroscopic orbits.}
\end{figure}

\section{Gaia's binary revolution and its limits}

The \emph{Gaia} mission has triggered a genuine revolution in binary star astrophysics~[3,4]. For the first time, a single all-sky experiment delivers astrometric, spectroscopic and photometric constraints on stellar multiplicity across a wide area of the Hertzsprung--Russell diagram. The DR3 Non-Single Star (NSS) catalogue contains more than $8\times 10^{5}$ astrometric, spectroscopic and eclipsing binary solutions~[5], and the dedicated variability pipelines provide over two million eclipsing binary candidates with light-curve models~[6]. These products have transformed our empirical view of nearby multiplicity, revealing structure in the period--eccentricity and mass-ratio distributions and enabling population studies of how binarity depends on stellar mass, age and environment~[7,8]. Forthcoming DR4/DR5 releases, with per-epoch radial velocities and longer astrometric baselines, will further expand the discovery space for spectroscopic and astrometric binaries~[4]. Despite this breakthrough, \emph{Gaia}'s view of stellar multiplicity is intrinsically incomplete and strongly shaped by selection effects. On the spectroscopic side, the RVS magnitude limit (full performance radial velocities for $G_{\rm RVS}\simeq 12$--13 and rapidly degrading precision towards $G_{\rm RVS}\sim 16$--17) excludes most low-mass M dwarfs, white dwarfs and many faint halo stars, precisely where binary fractions and period distributions are most uncertain. The scanning law and finite mission duration impose sparse, irregular sampling that favours periods of a few days to a few thousand days and disfavors both ultra short period binaries (hours to $<1$~day, where phase smearing and aliasing become severe) and long period systems with $P > 10^{4}$~days, where only accelerations or partial orbits are measured. Sensitivity to spectroscopic binaries further depends on inclination, mass ratio and semi-amplitude, so many low amplitude systems (small $K$) and highly eccentric orbits with poorly sampled periastron passages will lack robust constraints on $P$ and $e$. Astrometric and eclipsing solutions suffer analogous limitations in inclination, companion mass, distance and period, and are especially incomplete for compact-object companions. In practice, our simulations\footnote{We use a simplified forward model of Gaia DR5-like sampling and errors (toy, but calibrated to reproduce the main period and amplitude sensitivities of the RVS and astrometric pipelines) to illustrate where \emph{Gaia} will, and will not, deliver robust orbital solutions.} indicate that \emph{Gaia} DR5 will deliver robust single-lined spectroscopic orbits (i.e. with reliable $P$ and $e$) only over a restricted region of parameter space: binaries with periods of a few days to a few thousand days, moderate eccentricities, and RV semi-amplitudes comfortably above the single-epoch RVS noise. Several key binary populations are therefore poorly represented or essentially absent from the DR5 orbit catalogue: (i) very short-period systems (hours to $\sim 1$~day), where phase smearing and aliasing prevent robust SB1 solutions; (ii) very long-period binaries with $P > 10^{4}$~days, for which DR5 samples only a small fraction of the orbit; (iii) low-amplitude systems with $K < 2$--$3\,\sigma_{\rm RV}$; and (iv) many faint, hot, rapidly rotating or crowded sources that fall beyond the effective RVS performance limits. A dedicated ``Binarity Beyond \emph{Gaia}'' spectroscopic programme is therefore required to fill these gaps in the orbital parameter space and turn \emph{Gaia}'s heterogeneous binary detections into a truly global, bias correctable census of stellar multiplicity.

\section{A dedicated spectroscopic survey of binary stars}

The patterns in Fig.~\ref{fig:wst_galatea_airmass} show that  robust DR5 orbits will largely be confined to a limited ``sweet spot'' in $(P,e,K)$ and to comparatively bright stars within the effective RVS limit. A dedicated wide-field, multi-epoch spectroscopic survey is therefore needed to (i) fill the gaps in $(P,e,q)$ left by \emph{Gaia}, (ii) extend binary statistics to faint and distant populations, and (iii) provide a spectroscopic backbone for time-domain and multi-messenger studies of compact binaries. The key observational requirements are:
\begin{itemize}
    \item \textbf{Depth and populations:} observe several magnitudes below the \emph{Gaia} RVS limit to include intrinsically faint primaries and low-$q$ companions across all Galactic components.
    \item \textbf{RV precision:} achieve $< 1~\mathrm{km\,s^{-1}}$ for bright targets and a few $\mathrm{km\,s^{-1}}$ at the survey limit, enabling sensitivity to low-$K$ systems and useful constraints on $P$ and $e$.
    \item \textbf{Cadence:} obtain repeat visits spanning hours to years, with $\sim$10--20 epochs per target as a baseline and denser sampling for $P<1$~day candidates.
    \item \textbf{Wide-area multiplexing:} cover large sky areas with high multiplex to map multiplicity versus environment, position, and chemistry, leveraging \emph{Gaia} astrometry.
    \item \textbf{Spectral resolution:} use $R\sim5{,}000$--$10{,}000$ for survey RVs, plus an optional $R\sim20{,}000$--$25{,}000$ mode for detailed abundances of benchmarks and compact-object candidates.
\end{itemize}

The survey design should focus on the regions of parameter space where \emph{Gaia} is weakest. In the short-period regime, multi-epoch spectroscopy with minimal phase smearing can recover precise SB1/SB2 orbits for post-common-envelope binaries, contact systems, and ultra-compact white dwarf binaries that \emph{Gaia} will often classify only as variables or eclipsing systems without robust $P$ and $e$. At long periods, repeated high-precision RVs over multi-year baselines will convert \emph{Gaia} accelerations and partial astrometric orbits into fully constrained spectroscopic or joint solutions, capturing wide binaries and the outer components of hierarchical multiples. At low amplitudes, deeper spectroscopy and improved RV precision will open up the population of low-$q$, low-$K$ companions that lie below the effective RVS sensitivity in DR5 (right panel of Fig.~\ref{fig:wst_galatea_airmass}). A dedicated ``Binarity Beyond \emph{Gaia}'' survey would also be tightly integrated with other major facilities. Cross-matching to \emph{Gaia}, Rubin/LSST~[9] and \emph{Roman}~[10] would enable joint selection of eclipsing, transiting, and astrometric binary candidates, followed by spectroscopic monitoring to obtain full orbits and dynamical mass ratios. For compact binaries, multi-epoch RVs of known or candidate double-degenerate systems would deliver orbital solutions, systemic velocities, and chemical compositions essential for interpreting \emph{LISA} detections~[11] and constraining the formation channels and merger rates of gravitational-wave sources. Together, these elements define a focused survey concept that transforms \emph{Gaia}'s heterogeneous, selection-biased binary detections into a coherent, bias correctable map of stellar multiplicity across the Galaxy, while providing critical spectroscopic infrastructure for binary science in the era of precision astrometry, time-domain imaging, and gravitational-wave astronomy.

\section*{References}

{\small
[1] Raghavan D. et al., 2010, ApJS, 190, 1,
[2] Moe M., Di Stefano R., 2017, ApJS, 230, 15,
[3] Prusti T. et al., 2016, A\&A, 595, A1,
[4] Brown A.~G.~A. et al., 2021, A\&A, 649, A1,
[5] Gaia Collaboration, Arenou F. et al., 2023, A\&A, 674, A34,
[6] Gaia Collaboration, Eyer L. et al., 2023, A\&A, 674, A13,
[7] Arenou F. et al., 2022, A\&A, 657, A7,
[8] Jorissen A., 2022, A\&A, 662, A11,
[9] Ivezi\'c {\v Z}. et al., 2019, ApJ, 873, 111,
[10] Spergel D. et al., 2015, arXiv:1503.03757,
[11] Amaro-Seoane P. et al., 2017, arXiv:1702.00786
}

\end{document}